\documentclass[preprint,12pt]{elsarticle}

\usepackage{hyperref}
\usepackage{dirtytalk}
\usepackage{pifont}
\usepackage[T1]{fontenc}
\usepackage{lmodern}
\usepackage{caption}
\usepackage{enumitem}
\usepackage{adjustbox}
\usepackage{multirow}
\usepackage{tabularx}
\usepackage{tcolorbox}
\usepackage[table,usenames,dvipsnames]{xcolor}
\usepackage{amsmath}
\usepackage{booktabs}
\usepackage{tikz}

\newcommand{\dcircle}[1]{%
  \tikz[baseline=(c.base)]{
    \node[draw,circle,inner sep=1pt](c){#1};
  }%
}
\usepackage{textcomp} 

\newcommand{\circled}[1]{\textcircled{\footnotesize #1}}

\usepackage{subcaption}
\def\BibTeX{{\rm B\kern-.05em{\sc i\kern-.025em b}\kern-.08em
    T\kern-.1667em\lower.7ex\hbox{E}\kern-.125emX}}

\usepackage{csquotes}
\usepackage[mathlines,switch]{lineno}

\definecolor{codegray}{rgb}{0.5,0.5,0.5}
\definecolor{codegreen}{rgb}{0,0.6,0}
\definecolor{codepurple}{rgb}{0.58,0,0.82}
\definecolor{backcolour}{rgb}{0.95,0.95,0.92}

\usepackage{listings}

\usepackage{amssymb}

\usepackage{float}   
\floatstyle{ruled} 
\newfloat{listing}{tbp}{lop}
\floatname{listing}{Listing}

\usepackage{listings}

\lstdefinelanguage{Rego}{
  morekeywords={package, import, default, not, true, false, null},
  sensitive=true,
  comment=[l]{\#},
  morecomment=[s]{/*}{*/},
  morestring=[b]"
}

\lstdefinestyle{regostyle}{
  basicstyle=\ttfamily\footnotesize,
  keywordstyle=\color{blue!70!black}\bfseries,
  commentstyle=\color{gray!70},
  stringstyle=\color{green!50!black},
  numbers=left,
  numberstyle=\tiny\color{gray},
  stepnumber=1,
  numbersep=8pt,
  frame=lines,
  framerule=0.5pt,
  framesep=2mm,
  breaklines=true
}

\lstdefinelanguage{bash}{
  morekeywords={sudo,apt,opa,eval,cd,ls,cat,echo,export},
  sensitive=true,
  morecomment=[l]{\#},
  morestring=[b]",
  morestring=[b]'
}

\lstdefinestyle{termstyle}{
  basicstyle=\ttfamily\footnotesize,
  keywordstyle=\bfseries\color{blue!70!black},
  commentstyle=\itshape\color{gray!70},
  stringstyle=\color{green!50!black},
  numbers=left,
  numberstyle=\tiny\color{gray},
  frame=lines,
  framerule=0.5pt,
  framesep=2mm,
  breaklines=true,
  showstringspaces=false,
  columns=fullflexible
}
\begin{document}

\begin{frontmatter}



\title{An Empirical Study of Policy-as-Code Adoption in Open-Source Software Projects}


\author{Patrick Loic Foalem\corref{c1}}

\ead{\{patrick-loic.foalem, foutse.khomh, leuson-mario-pedro.da-silva, ettore.merlo\}@polymtl.ca} 
\affiliation{organization={Department of Computer and Software Engineering},
            addressline={Polytechnique Montreal}, 
            city={Montreal},
            postcode={H3T 1J4},
            state={Quebec},
            country={Canada}}

\author{Foutse Khomh}
\author{Leuson Da Silva}
\author{Ettore Merlo}
\cortext[c1]{Corresponding author.}


\begin{abstract}
\textbf{Context:} Policy-as-Code (PaC) has become a foundational approach for embedding governance, compliance, and security requirements directly into software systems. While organizations increasingly adopt PaC tools, the software engineering community lacks an empirical understanding of how these tools are used in real-world development practices.

\textbf{Objective:} This paper aims to bridge this gap by conducting the first large-scale study of PaC usage in open-source software. Our goal is to characterize how PaC tools are adopted, what purposes they serve, and what governance activities they support across diverse software ecosystems.

\textbf{Method:} We analyzed 399 GitHub repositories using nine widely adopted PaC tools. Our mixed-methods approach combines quantitative analysis of tool usage and project characteristics with a qualitative investigation of policy files. We further employ a Large Language Model (LLM)--assisted classification pipeline, refined through expert validation, to derive a taxonomy of PaC usage consisting of 5 categories and 15 sub-categories.

\textbf{Results:} Our study reveals substantial diversity in PaC adoption. PaC tools are frequently used in early-stage projects and are heavily oriented toward governance, configuration control, and documentation. We also observe emerging PaC usage in MLOps pipelines and strong co-usage patterns, such as between OPA and Gatekeeper. Our taxonomy highlights recurring governance intents.

\textbf{Conclusion:} Our findings offer actionable insights for practitioners and tool developers. They highlight concrete usage patterns, emphasize actual PaC usage, and motivate opportunities for improving tool interoperability. This study lays the empirical foundation for future research on PaC practices and their role in ensuring trustworthy, compliant software systems.
\end{abstract}


\end{frontmatter}

\section{INTRODUCTION}
\label{sec:introduction}

Modern software systems operate in hyper-connected environments, continuously collecting and processing vast amounts of data. As a result, they are increasingly exposed to security violations, data leaks, and operational failures~\cite{7961663}, which can lead to privacy breaches, regulatory non-compliance, financial penalties, service interruptions, and long-term loss of trust~\cite{anwar2018understanding}. To address these challenges, requirements engineering and regulatory bodies have emphasized the role of policies in ensuring security, compliance, traceability, and operational governance~\cite{taylor1994guidelines}. Historically, policies were manually enforced or hard-coded, making them difficult to evolve, audit, or verify. With the rise of dynamic, distributed, and cloud-native systems, governance mandates such as GDPR, HIPAA, and the EU AI Act~\cite{act2024eu} have created demand for scalable, automated compliance mechanisms throughout the software lifecycle.  
\textit{Policy as Code} (PaC) addresses those previously mentioned needs by encoding policies as machine-readable, version-controlled, and executable artifacts. By embedding policy logic into infrastructure and software pipelines, PaC enables proactive enforcement of security, compliance, and operational rules, thereby reducing risks of misconfiguration, unauthorized access, and regulatory violations. This approach enhances automation, traceability, and integration with modern practices such as DevOps and CI/CD~\cite{pac3}. A growing ecosystem of PaC tools supports this paradigm, including Open Policy Agent (OPA), Kyverno, Cloud Custodian, and HashiCorp Sentinel. OPA is a general-purpose engine with the Rego language, often integrated into platforms such as Kubernetes and Terraform~\cite{pac10}, while Kyverno is a Kubernetes-native engine that simplifies policy authoring through YAML configurations~\cite{pac11}.  
Despite these advances, to the best of our knowledge, there is little large-scale empirical evidence on how PaC tools are adopted and used in open-source software (OSS). Understanding their adoption is essential for identifying gaps in developer practices, guiding tool improvement, and informing future research on policy-driven governance.  
As a result, this paper presents the first large-scale empirical study of PaC tool adoption in OSS. We curate a comprehensive list of tools and mine a dataset of 399 GitHub repositories that employ at least one PaC tool. Our study is driven by the following research questions:

\begin{itemize}
  \item \textbf{RQ1: What is the popularity of different PaC tools in OSS projects?} We analyze the usage of PaC tools across OSS repositories and find that OPA, Cloud Custodian, Kyverno, and Pulumi are the most popular tools. 
  While our findings can guide practitioners toward widely adopted tools with strong community support, widely used tools provide researchers with a baseline for studying adoption trends and practitioners with guidance toward stable, well-supported solutions.


  \item \textbf{RQ2: What kind of projects adopt PaC tools?} Through manual analysis, we observe that PaC tools are primarily adopted in projects, that can be classified into categories such as \textit{documentation (37\%)}, \textit{tooling (32\%)}, and \textit{DevOps infrastructure (17\%)}. 
  We also find emerging adoption in \textit{MLOps (1\%)} and \textit{AI/Research (0.2\%)}, indicating a growing interest in using PaC tools to enforce governance within intelligent systems. These findings can guide developers in understanding the types of projects where PaC tools are commonly used, helping them envision appropriate adoption scenarios.

  \item \textbf{RQ3: How is PaC used in practice across different tools?} We manually inspected policy files to understand their practical purposes and developed a structured taxonomy of PaC usage, comprising five high-level categories: 
  \textit{Workflow Automation}, 
  \textit{Cost Optimization}, \textit{Security}, \textit{Compliance}, and \textit{Deployment Governance}. Understanding the purpose of usage helps practitioners in their adoption process, as they can apply these tools for similar objectives, while also assisting tool developers in designing more effective tools that integrate most of these use cases. 
  

  \item \textbf{RQ4: Are different PaC tools used together?} Understanding whether PaC tools are used in combination is essential for assessing how organizations manage interoperability, policy consistency, and governance complexity across their infrastructure. To explore this, we examine co-usage patterns across projects and tools. Our analysis reveals that while most projects tend to adopt a single PaC tool, OPA frequently appears in combination with other tools. These insights can support practitioners in designing multi-tool adoption strategies and guide tool builders toward improved standardization and interoperability support.

\end{itemize}


\textbf{Paper Organization:} The remainder of the paper is organized as follows. Section~\ref{sec:background} introduces key background concepts on policies in software engineering and the PaC paradigm. Section~\ref{sec:approach} details our study methodology. Section~\ref{sec:result} presents our results and answers to the research questions.  Section~\ref{sec:discussion_implication} discusses the implications of our findings. 
In Section~\ref{sec:related_work}, we position our study within the context of related work.
Section~\ref{sec:threats_to_validity} outlines threats to validity. We conclude and propose future research directions in Section~\ref{sec:conclusion}.

\section{BACKGROUND}
\label{sec:background}
To investigate the usage of PaC tools in modern software systems, we first clarify the core concepts and typical system architecture. In this paper, we use \texttt{PaC tools} to denote libraries or frameworks whose primary purpose is to express and enforce policies through code-based mechanisms.  

\textbf{Policy as Code (PaC).}  
PaC specifies and enforces policies as executable rules evaluated by a dedicated policy engine. A policy is a set of conditions governing system behavior, where the engine returns outcomes such as \emph{allow}, \emph{warn}, or \emph{deny}. By exposing decisions through APIs and keeping policies separate from application code, PaC decouples governance from business logic and enables consistent, reusable, and testable enforcement across services and environments~\cite{pac24}.  

\begin{figure}[H]
    \centering
    \includegraphics[width=0.4\textwidth]{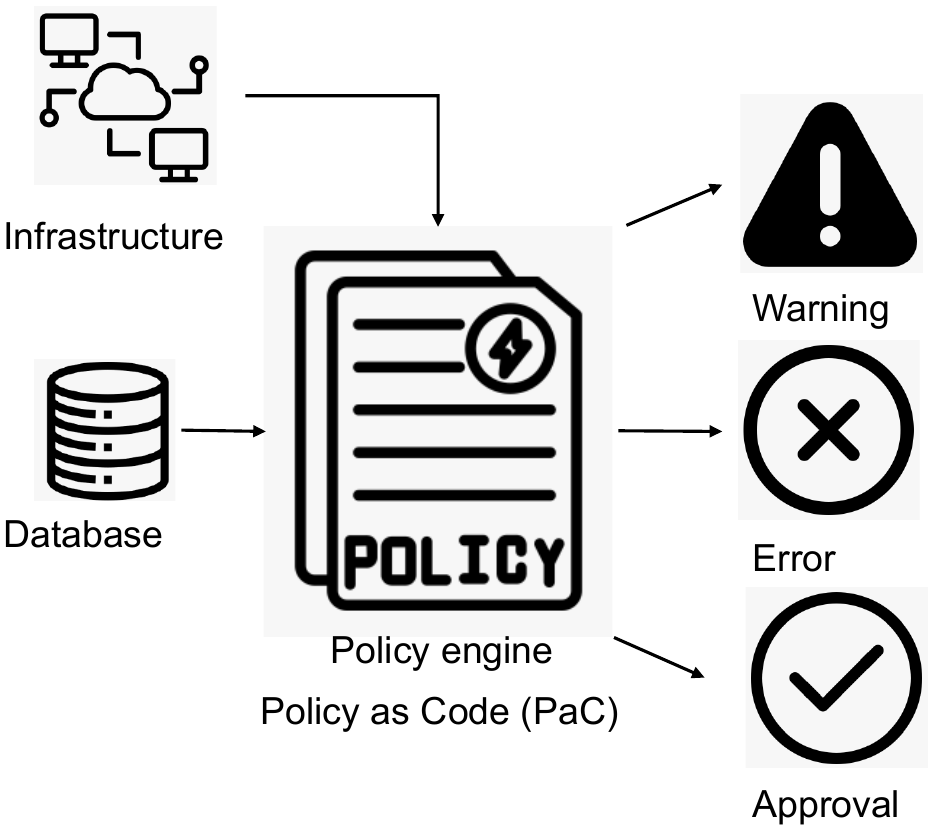}
    \caption{Policy-as-Code architecture: policies are authored once and enforced consistently by a centralized engine across services and pipelines.}
    \label{fig:architecture}
\end{figure}

\autoref{fig:architecture} illustrates a reference PaC architecture. Policies are typically written in declarative formats (e.g., JSON or YAML) or a domain-specific language (Listing~\ref{lst:rego-policy}) processed by the engine. The engine consumes contextual inputs (e.g., request parameters, user identity, or infrastructure metadata), evaluates the rules, and returns a decision with optional explanations. This design supports multiple integration modes--embedded library, sidecar, admission controller, or standalone service--making PaC suitable for CI/CD pipelines and Infrastructure-as-Code (IaC) workflows~\cite{pac24}.  

As an example, consider an organization that requires all AWS EC2 instances provisioned via Terraform to include an \texttt{Env} tag for auditing and cost tracking. With OPA, this requirement can be expressed as a Rego rule that inspects a Terraform plan and flags any instance missing the tag (Listing~\ref{lst:rego-policy}). The policy is evaluated using the OPA CLI (Listing~\ref{lst:opa-eval}), which loads the policy and the plan as input and reports violations. This centralizes policy logic, ensures uniform enforcement at deploy time, and reduces manual review effort in continuous delivery settings~\cite{pac24}.

\begin{listing}[H]
\caption{OPA policy to enforce \texttt{Env} tag on EC2 instances.}
\label{lst:rego-policy}
\begin{lstlisting}[language=Rego,style=regostyle]
package ec2

deny[msg] {
  resource := input.resource_changes[_]
  resource.type == "aws_instance"
  not resource.change.after.tags.Env

  msg := sprintf("EC2 instance %v is missing required 'Env' tag", [resource.address])
}
\end{lstlisting}
\end{listing}

\begin{listing}[H]
\caption{Evaluate the OPA deny rule for EC2.}
\label{lst:opa-eval}
\begin{lstlisting}[language=bash,style=termstyle]
opa eval --data policy.rego --input plan.json "data.ec2.deny"
\end{lstlisting}
\end{listing}

\section{STUDY DESIGN}
\label{sec:approach}
\subsection*{\textbf{\circled{1} Identifying PaC Tools from Gray Literature}}
To identify PaC tools suitable for our study, we performed a search on Google \footnote{\url{https://www.google.ca/}} using the query: \texttt{``(``policies as code'' OR ``policy as code'' OR PAC) AND (tools OR solutions OR frameworks)''}. We relied solely on Google because, to the best of our knowledge, no prior academic studies have systematically catalogued PaC tools. This search was conducted in March 2025. Following the methodology of similar prior studies \cite{foalem2025logging, foalem2024studying}, we collected all relevant PaC tools mentioned within the first 100 search results. In total, we identified 28 unique PaC tools. Which can be found in our replication package \cite{replication}. 

\subsection*{\textbf{\circled{2} Filtering PaC Tools}}
The first two authors jointly reviewed the official documentation of each tool during a series of discussion meetings to apply our inclusion criteria. Tools were retained if they were (i) open-source and (ii) primarily designed for implementing policies as code, or if they provided clear documentation on PaC-specific modules, languages, or enforcement mechanisms. Importantly, we focused on tools whose core purpose is PaC, rather than multi-purpose tools that blend policy features with broader configuration or provisioning capabilities. This choice avoids conceptual overlap and prevents conflating Infrastructure-as-Code (IaC) functionalities with dedicated policy specification and enforcement mechanisms.

For instance, although tools such as Terraform, Chef, and Ansible are occasionally cited as PaC tools in some sources \cite{pac1}, their primary purpose—according to their official documentation—is Infrastructure as Code (IaC), making it difficult to systematically distinguish IaC components from PaC components. After applying these criteria, we retained 9 PaC tools for our study. The complete list is available in \autoref{tab:pac_tools_overview}.
 

\begin{table*}[htbp]
\centering
\caption{Overview of PaC Tools with Metadata, Language, and Descriptions. Size in KB.}
\begin{adjustbox}{width=1\textwidth}
\begin{tabular}{
    >{\raggedright\arraybackslash}p{3.2cm}   
    >{\raggedright\arraybackslash}p{1.6cm}   
    >{\raggedright\arraybackslash}p{1.6cm}   
    >{\raggedright\arraybackslash}p{1.6cm}   
    >{\raggedright\arraybackslash}p{1.6cm}   
    >{\raggedright\arraybackslash}p{1.8cm}   
    >{\raggedright\arraybackslash}p{7.6cm}   
}
\toprule
\rowcolor{gray!15}
\textbf{PaC Tool} & \textbf{Size (KB)} & \textbf{Stars} & \textbf{Forks} & \textbf{Releases} & \textbf{Language} & \textbf{Description} \\
\midrule

Open Policy Agent (OPA) & 1,091,526 & 10,338 & 1,425 & 117 & Go & An open source, general-purpose policy engine~\cite{pac15}. \\
\rowcolor{gray!5}

HashiCorp Sentinel & 330 & 52 & 13 & 10 & Go & A logic-based policies language for managing your policy.~\cite{pac16}. \\
\rowcolor{gray!15}

Pulumi & 187,114 & 23,247 & 1,204 & 313 & Go & A crossGuard policy-as-code for IaC and account insights~\cite{pac17}. \\
\rowcolor{gray!5}

Cedar Policy Language & 8,582 & 1,041 & 100 & 38 & Rust & A language for writing policies in your applications~\cite{pac18}. \\
\rowcolor{gray!15}

Kyverno OSS & 151,089 & 6,414 & 1,028 & 250 & Go & A cloud-native Kubernetes policy management~\cite{pac19}. \\
\rowcolor{gray!5}

Cloud Custodian & 136,109 & 5,694 & 1,545 & 78 & Python & A rules engine for managing policies in cloud infrastructure~\cite{pac20}. \\
\rowcolor{gray!15}

AWS Config & 4,595 & 468 & 172 & 38 & Python & A tool that allows developers to write policies rules in AWS.~\cite{pac21}. \\
\rowcolor{gray!5}

GateKeeper & 182,832 & 3,909 & 793 & 121 & Go & A kubernetes policy controller~\cite{pac22}. \\
\rowcolor{gray!15}

Kubewarden & 2,982 & 209 & 37 & 122 & Go & A dynamic admission controller that uses policies~\cite{pac23}. \\
\bottomrule
\end{tabular}
\end{adjustbox}
\label{tab:pac_tools_overview}
\end{table*}

\subsection*{\textbf{\circled{3} Collecting GitHub Projects Using PaC Tools}}

\begin{figure}
    \centering
    \includegraphics[width=0.8\textwidth]{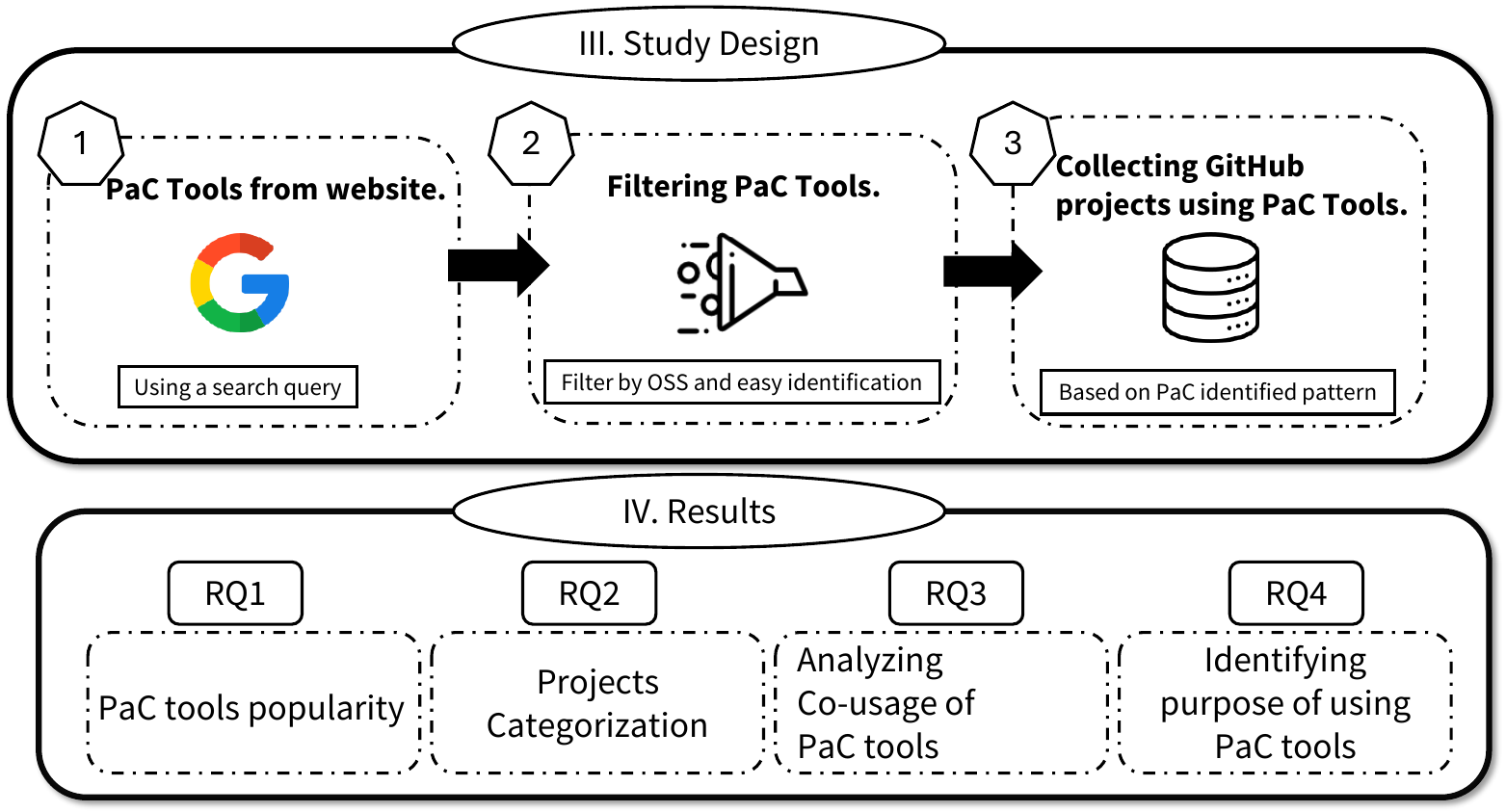}
    \caption{Overview of our research workflow.}
    \label{fig:research-workflow}
\end{figure}

Once we established the set of PaC tools, we moved into our third step (\autoref{fig:research-workflow}).
For that, we used the GitHub Search API \cite{pac2} to collect open-source projects that use the selected PaC tools. Based on the official documentation of each tool, we identified unique patterns that characterize the adoption of each PaC tool within a GitHub repository. 
For instance, OPA and HashiCorp Sentinel use specific languages that can be identified through file extensions (\texttt{.rego} and \texttt{.sentinel}, respectively), while other tools like Gatekeeper can be detected by searching for YAML files containing the keyword \texttt{ConstraintTemplate}. 
The complete list of detection patterns is provided in our replication package \cite{replication}.

This collection process initially yielded 6,010 candidate repositories for all PaC tools. 
We then removed 68 duplicate entries, since the search was at the file level, resulting in 5,942 unique repositories. 
To filter out irrelevant or outdated projects, we applied the following filtering criteria, adapted from best research practices \cite{foalem2025logging, openja2022studying, gonzalez2020state}:

\begin{itemize}
  \item \textbf{Size:} Repository must have a size $>$ 0 KB.
  \item \textbf{Popularity:} Repository must have $\geq$ 5 stars \textit{or} and $\geq$ 5 forks.
  \item \textbf{Activity:} The last commit must have been within 2024.
\end{itemize}

\begin{figure}
    \centering
    \includegraphics[width=0.6\textwidth]{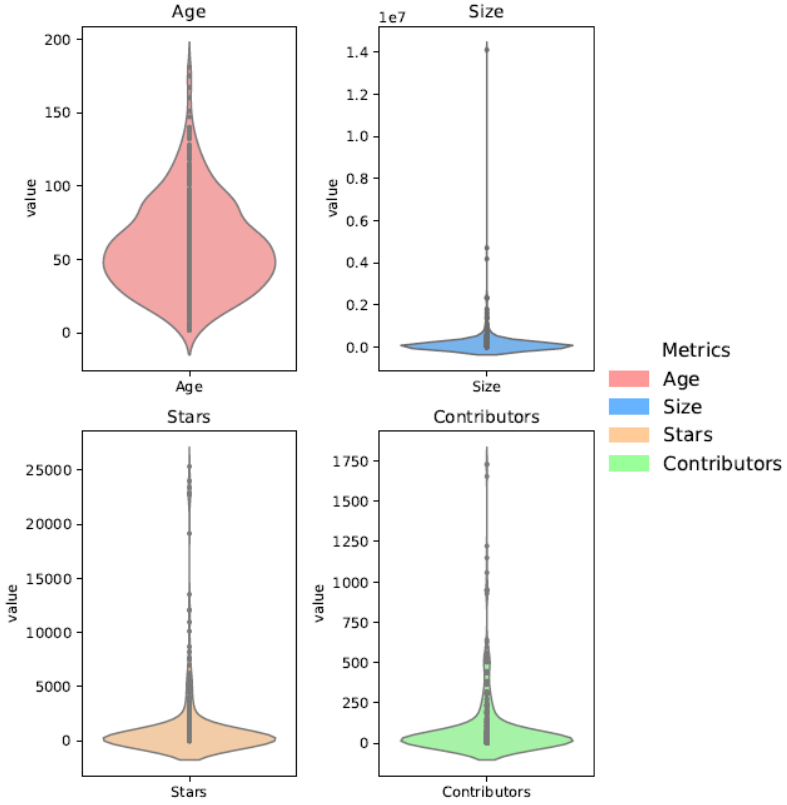}
    \caption{Distribution of project metrics using PaC tools: Age (months), Size (bytes), Stars (popularity), and Contributors (collaboration) across our study projects.}
    \label{fig:Overview}
\end{figure}

After applying these filtering criteria, we retained a final dataset of 399 OSS projects for analysis.
Figure~\ref{fig:Overview} shows that approximately 75\% of the remaining projects have low to moderate values across key metadata attributes (e.g., fewer than 10 contributors, 100 stars, 50 months of age, or 1MB in size), while a few projects stand out as mature or highly active.
Data analysis based on this project sample is presented in Section~\ref{sec:result}.


\section{RESULTS}
\label{sec:result}
\subsection{\textbf{RQ1: What is the popularity of different PaC tools in OSS projects?}}

\subsubsection{\textbf{Motivation}} 
PaC tools are designed to define and enforce policies through high-level programming languages~\cite{pac3}. These tools enable organizations to automate policy enforcement across cloud-native environments as part of the software development lifecycle~\cite{pac3}. However, integrating PaC tools into software workflows can be challenging due to the diversity of available tools and the complexity of their configurations~\cite{pac4, pac5, pac6}. Investigating the popularity of these tools in OSS projects provides valuable insight into which tools are most widely adopted and supported in practice. This information can guide cloud-native development teams in selecting PaC tools that are not only technically capable but also backed by active communities, better documentation, and practical usage examples. The popularity of a tool is often a proxy for its maturity and reliability, making it a critical factor in adoption decisions. 
Furthermore, they represent a baseline for researchers for studying adoption trends and practitioners with guidance toward stable, well-supported solutions.

\subsubsection{\textbf{Approach}} 
To analyze the popularity of PaC tools across the 399 projects using them, we relied on two complementary metrics:

\begin{itemize}
    \item \textbf{Number of files containing PaC code:} Since some PaC tools are built around domain-specific languages (e.g., \texttt{.rego}, \texttt{.sentinel}, \texttt{.cedar}), while others are implemented within general-purpose languages (e.g., \texttt{Python}, \texttt{Java}), we opted for file-level analysis rather than line of code level granularity to minimize measurement bias~\cite{openja2022studying}. Based on the identified patterns defined in Section~\ref{sec:approach}, we scanned all repositories of our sample, counting the number of files matching each tool's pattern. A higher number of matched files indicates greater tool prevalence within projects.
    
    \item \textbf{Number of repositories using a PaC tool:}  To complement the previous metric, we introduce a second one, the number of distinct repositories that include at least one file associated with a given PaC tool. This metric is essential to assess the distribution and popularity of tools across diverse projects. For instance, a tool might be heavily used in a few large projects (high file count, low repo count) or lightly used across many projects (low file count, high repo count). A repository is considered to use a PaC tool if it contains at least one file corresponding to one of the tool's identified patterns. Therefore, a higher number of repositories using a given tool reflects broader adoption across the OSS ecosystem. 
\end{itemize}

Together, these two metrics offer a better perspective on the popularity of PaC tools by capturing both their depth of use within individual projects and their breadth of adoption across projects.





\subsubsection{\textbf{Result}}

\begin{figure}[h]
    \centering
    
    \begin{subfigure}{\textwidth}
        \centering
        \includegraphics[width=0.6\textwidth]{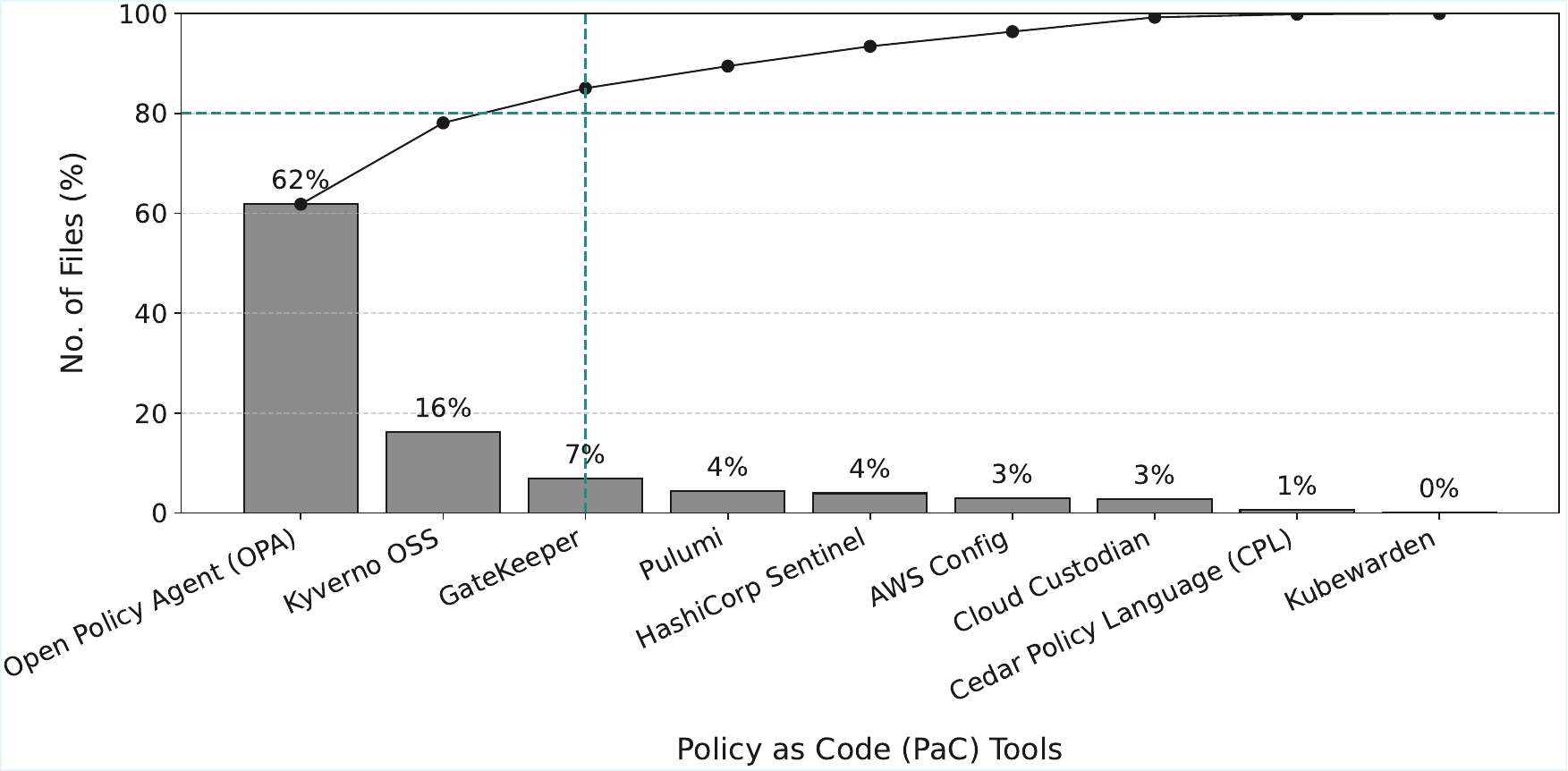}
        \caption{Percentage of PaC usage based on file counts.}
        \label{fig:pac_usage_file}
    \end{subfigure}
    
    \vspace{1em} 

    \begin{subfigure}{\textwidth}
        \centering
        \includegraphics[width=0.6\textwidth]{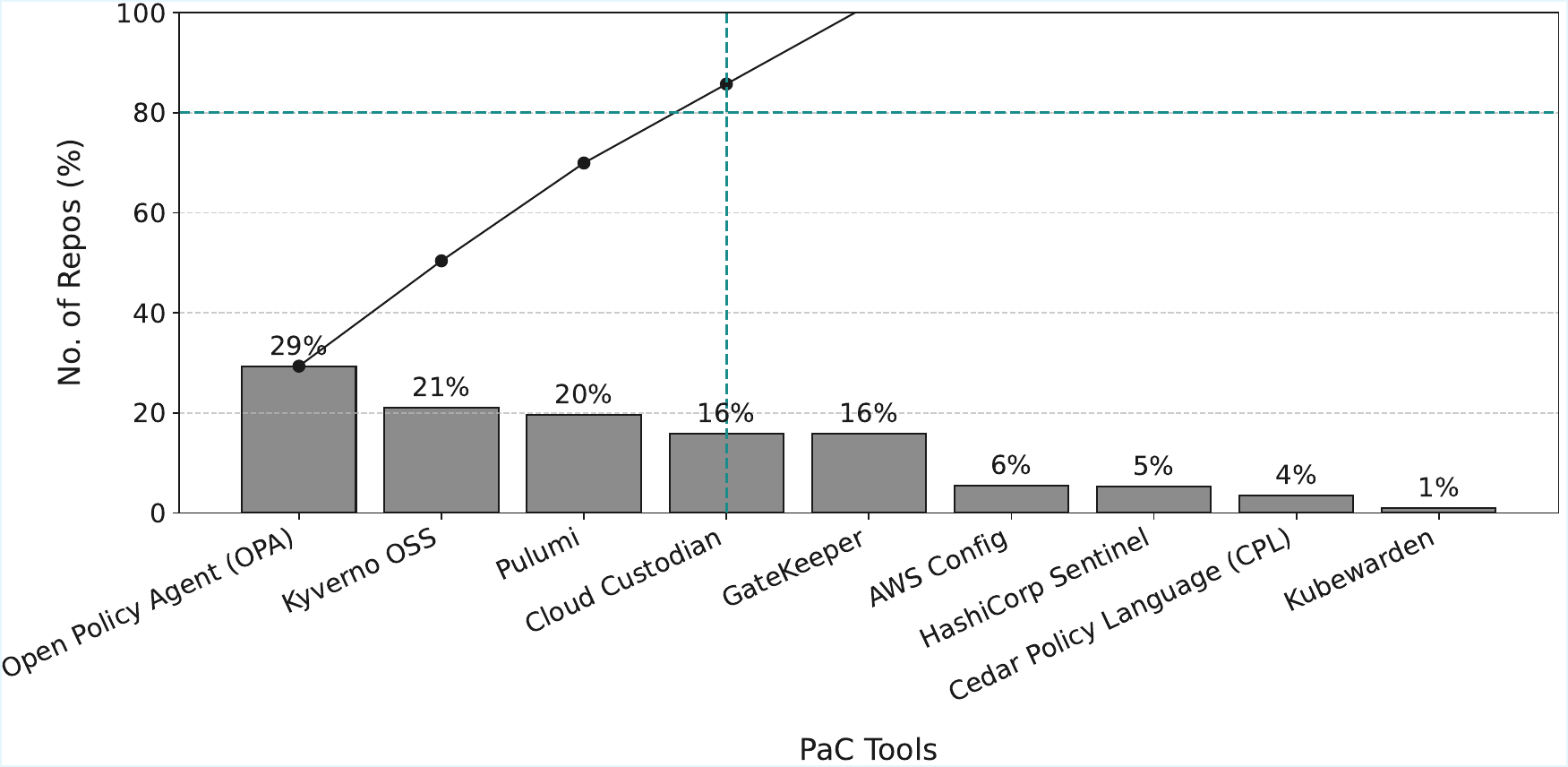}
        \caption{Percentage of PaC usage based on the number of repositories.}
        \label{fig:pac_usage_repo}
    \end{subfigure}
    
    \caption{Comparison of Policy as Code (PaC) tool usage}
    \label{fig:pac_usage_combined}
\end{figure}

\autoref{fig:pac_usage_combined} summarizes the popularity of the nine PaC tools across our dataset of 399 projects.
\autoref{fig:pac_usage_file} shows the percentage of files matching PaC tool patterns, capturing the intensity of tool usage within projects, while \autoref{fig:pac_usage_repo} reports the percentage of repositories using at least one file from a given PaC tool, reflecting adoption breadth.

From the file-based analysis (\autoref{fig:pac_usage_file}), OPA dominates PaC usage, accounting for 62\% of all policy-related files that rely on OPA, followed at a distance by Kyverno (16\%) and GateKeeper (7\%). 
Pulumi and HashiCorp Sentinel each account for 4\%, while AWS Config and Cloud Custodian contribute 3\% each. Finally, the Cedar Policy Language (1\%) shows lower but still notable adoption, whereas Kubewarden (<1\%) has only a minimal presence. 
Finally, the top three tools--OPA, Kyverno, and GateKeeper--account for 85\% of the total PaC-related file count, illustrating their central role in active policy implementation.

In contrast, the repository-level analysis (\autoref{fig:pac_usage_repo}) shows a more balanced distribution. 
OPA appears in 29\% of projects, followed by Kyverno (21\%), Pulumi (20\%), Cloud Custodian (16\%), and GateKeeper (16\%). AWS Config (6\%), HashiCorp Sentinel (5\%), the Cedar Policy Language (4\%), and Kubewarden (1\%) round out the list. Collectively, the top four tools--OPA, Kyverno, Pulumi, and Cloud Custodian--account for 82\% of the repositories.

Overall, while OPA is the most deeply used tool (in terms of file count), other tools--especially Kyverno, Pulumi, and Custodian--demonstrate competitive breadth of adoption across OSS projects. 

\begin{tcolorbox}[colback=gray!5!white,colframe=gray!75!black]
All nine analyzed PaC tools are actively used by developers during the software development lifecycle. Among them, \textbf{OPA}, \textbf{Kyverno}, \textbf{Pulumi}, and \textbf{Cloud Custodian} emerge as the top four most widely adopted tools across projects.
\end{tcolorbox}

\subsection{\textbf{RQ2: What kind of projects adopt PaC tools?}}
\subsubsection{\textbf{Motivation}} 
Identifying the categories of projects that use PaC tools can reveal practical contexts where PaC provides value. Understanding this landscape is critical for several reasons. First, it informs tool developers about the domains and development environments where their tools are being used, supporting future improvements and integration strategies. Second, it helps practitioners make informed decisions about where to adopt PaC or enhance existing development practices by aligning with project categories where PaC is commonly used, thereby supporting the development of more secure and governable software systems. Finally, it offers researchers a foundation for further empirical studies into the effectiveness, challenges, and potential of PaC in specific software development contexts.

\subsubsection{\textbf{Approach}} 
To answer this research question, we conducted a manual qualitative analysis of the open-source repositories that use PaC tools. This analysis was carried out by two researchers with strong backgrounds in Software Engineering for AI (SE4AI) and AI for Software Engineering (AI4SE).

First, our labeling process began with a group discussion on an initial set of 20 repositories. The objective was to review these samples together and establish a shared understanding of the labeling categories. For this, we used the project taxonomy proposed by Majidi et al.~\cite{majidi2022empirical} as a baseline and adapted it based on empirical evidence from our dataset, while introducing new categories when applicable.
During labeling, we examined a variety of project metadata, including the GitHub repository name and topics, project description, README content, and, in some cases, the actual source code when additional context was needed. The time required to label a single project varied between 45 seconds and 30 minutes, depending on the complexity and clarity of the project documentation.

Moving on, in order to validate and improve our understanding of the categories and their definitions, each researcher independently labeled 40\% of our sample of repositories (162 repositories). Next, we calculated the Cohen's Kappa agreement score ($\kappa$), which yielded a substantial agreement score of 0.77, indicating good consistency between the raters. 
All possible disagreements were discussed and resolved collaboratively, while also contributing to the refinement of our target categories.
Following this harmonization step, we proceeded to label the remaining 217 repositories in parallel, applying the refined category schema. 
This phase achieved a near-perfect agreement score of 0.81. 
Once again, any remaining disagreements were resolved via collaborative discussion.
As a result of this iterative manual inspection and reconciliation process, we categorized the dataset into six distinct project categories: DevOps, Toolkit, MLOps, Documentation, AI/Research, and Application System.

\subsubsection{\textbf{Result}} 

\begin{table*}[htbp]
\centering
\caption{Distribution of Project Categories with Descriptions and Median Repository Metrics. 
The columns report the number and proportion of repositories per category, along with the median values for repository size (in kilobytes), number of contributors, forks, and stars.}
\begin{adjustbox}{width=0.9\textwidth}
\begin{tabular}{
    >{\raggedright\arraybackslash}p{3.2cm}  
    >{\raggedright\arraybackslash}p{7.5cm}  
    >{\centering\arraybackslash}p{1.5cm}    
    >{\centering\arraybackslash}p{1.5cm}    
    >{\centering\arraybackslash}p{1.5cm}    
    >{\centering\arraybackslash}p{1.5cm}    
    >{\centering\arraybackslash}p{1.5cm}    
}
\toprule
\rowcolor{gray!15}
\textbf{Category} & \textbf{Descriptions} & \textbf{Repos (\%)} & \textbf{Size (KB)} & \textbf{Contrib} & \textbf{Forks} & \textbf{Stars} \\
\midrule

DevOps & Projects that leverage infrastructure and operations tools (e.g., Kubernetes, Ansible, Docker) to automate the provisioning, deployment, and governance of software applications. & 68 (17\%) & 49,473 & 59 & 88 & 329 \\
\addlinespace

Toolkit & Standalone libraries, frameworks, APIs, plugins, or modules that offer reusable functionalities or components to simplify software development. & 127 (32\%) & 13,354 & 23 & 40 & 152 \\
\addlinespace

MLOps & These projects combine AI/ML models with DevOps tools and practices to automate the ML lifecycle, including model training, deployment, monitoring, and governance. & 6 (1\%) & 10,104 & 4 & 33 & 164 \\
\addlinespace

Documentation & Projects that primarily serve as documentation, tutorials, workshops, demo or use case. & 147 (37\%) & 11,735 & 15 & 29 & 46 \\
\addlinespace

AI/Research & Academic or experimental projects involving AI/ML models or techniques. & 1 (0.2\%) & 200 & 3 & 6 & 43 \\
\addlinespace

Application System & These are software projects or programs. They may include web applications or traditional systems without AI/ML components. & 50 (12\%) & 26,253 & 22 & 31 & 80 \\
\midrule
\rowcolor{gray!10}
\textbf{Total} &  & \textbf{399 (100\%)} & \textbf{111,119} & \textbf{126} & \textbf{227} & \textbf{814} \\
\bottomrule
\end{tabular}
\end{adjustbox}
\label{tab:project_categories_summary}
\end{table*}

Our manual categorization of our sample of GitHub repositories (399) shows that PaC tools are adopted across a diverse range of project types.
\autoref{tab:project_categories_summary} presents the distinct categories we identified (6) along with their median repository characteristics.
The most prevalent category is \textbf{Documentation} (37\%), comprising tutorials, workshops, and example use cases (e.g., \textit{aws-samples/cfn101-workshop}~\cite{repo1}).
These projects tend to be modest in size and activity, with a median of 15 contributors, reflecting their role as educational or demonstrative resources. 
Next, we have the category \textbf{Toolkit} (32\%), mainly consisting of reusable libraries, APIs, and modules supporting software development (e.g., \texttt{kubescape/regolibrary}~\cite{repo2}). These projects show moderate engagement, enabling integration and extension of PaC capabilities. 

\textbf{DevOps} projects  (17\%) adopt PaC tools in infrastructure automation and governance workflows, consistent with their intended use in cloud-native systems (e.g., \texttt{cloudposse/atmos}~\cite{repo3}). These repositories show the highest median values across all metrics, indicating production-grade adoption.
\textbf{Application System (12\%)} includes software systems--often web-based--that embed PaC tools to enforce policies at runtime (e.g., \texttt{aquasecurity/postee}~\cite{repo4}).
\textbf{MLOps (1\%)} projects, though few, integrate PaC tools within ML pipelines to support secure, trustworthy AI practices (e.g., \texttt{deployKF/deployKF}~\cite{repo5}). Their presence signals early but growing interest in applying policy governance to AI systems.
Finally, we identified a single project under the \textit{AI/Research} category (\texttt{DataBassGit/AssistAF}~\cite{repo6}). 

\begin{tcolorbox}[colback=gray!5!white,colframe=gray!75!black]
Overall, we observed that PaC tools are predominantly used in \textit{documentation} and \textit{toolkits}, reflecting their current ecosystem-building phase. 
Nonetheless, their adoption extends to a diverse range of domains, including \textit{DevOps}, \textit{application systems}, and \textit{MLOps}. 
\end{tcolorbox}

\subsection{\textbf{RQ3: How is PaC used in practice across different projects?}}   
\label{sec:RQ3_result}
\subsubsection{\textbf{Motivation}} 
While prior questions establish the popularity and adoption contexts of PaC tools, they do not inform the types of policies being implemented. 
Understanding how PaC is used in practice--i.e., the specific policy concerns being encoded--provides deeper insights into the functional roles these tools play within real-world software projects. 
Identifying which types of policies are most commonly defined can help assess how these tools align with developer intentions and organizational priorities.
Answering this question offers practical value for tool developers, as (i) it helps to identify which policy concerns receive the most attention, where existing tools may fall short, and (ii) how they can be improved to better align with developer workflows and organizational goals.

\subsubsection{\textbf{Approach}} 
To investigate how PaC tools are used in practice and to identify the target resources governed by these policies, we employed a stratified random sampling approach combined with expert-guided policy analysis~\cite{abbassi2025unveiling, ouni2023empirical, chen2020comprehensive}.
For that, we followed the next steps:


\subsubsection*{\dcircle{1} Data Collection}
From the full dataset (12,152 files from 399 repositories), we extracted a total of 373 PaC policy files using a 95\% confidence level and a 5\% confidence interval~\cite{foalem2024studying}. Stratified random sampling ensured proportional representation across different PaC tools.

\subsubsection*{\dcircle{2} LLM-as-a-Judge - PaC Analysis} 
To analyze and classify policies, we adopt the \textit{LLM-as-a-Judge} approach, which treats LLMs as reference-free evaluators capable of  categorizing inputs based on structured schema~\cite{11121731}. This approach has shown promise in various domains of software engineering, including code summarization, bug triaging, and automated documentation~\cite{abbassi2025unveiling, 11121731}. Inspired by this framework, each policy from the selected subset of files was analyzed using a LLM, specifically \texttt{gpt-4o-mini}, configured with a temperature of 0.2~\cite{abbassi2025unveiling}. 
The LLM was prompted with a detailed schema to classify each policy across seven structured dimensions: Primary Purpose, Sub-purposes, Taxonomy Category, Taxonomy Sub-category, Policy Implemented, and Rationale. The complete prompt and logic are included in our replication package~\cite{replication}. The prompt guided the LLM to provide its answer in JSON.
\subsubsection*{\dcircle{3} Manual Labelling - Pilot Study} 
Knowing that no previous study defined categories for PaC, our goal is to build a taxonomy from our set of selected policies.
For that, we initially rely on the judgment of LLMs. Our approach is inspired by recent empirical work that also used LLM-generated judgments as a starting point for taxonomy construction~\cite{abbassi2025unveiling}, which developed a taxonomy of inefficient Python code generated by LLM using LLM-assisted analysis. Following this methodological precedent, we used LLM outputs as an initial guide. 
To start this process, we randomly selected a sample of 111 (\( S \)), representing 30\% of the total sample \( T \), while maintaining proportional representation across all identified PaC tools. 
To ensure quality and reliability, two researchers with expertise in Software Engineering and Machine Learning collaboratively reviewed and refined the LLM-generated taxonomy entries through discussion. 
Using an open coding approach~\cite{seaman1999qualitative}, the researchers systematically evaluated each policy file by examining the LLM-assigned category and sub-category in relation to the content and structure of the policy. During this process, several issues with the raw LLM labels emerged. The LLM frequently produced overly specific categories such as \emph{``Kubernetes Workloads''}, \emph{``Resource Paths''}, \emph{``Network Access''}, or \emph{``Infrastructure Deployment Configurations''}. Through open coding, these fragmented or overly narrow labels were abstracted into broader, categories such as \emph{Workload Management}, \emph{``Resource Management''}, and \emph{``Network Management''}.
Similarly, the LLM often used semantically overlapping high-level categories—such as \emph{``Security Governance''}, \emph{``Access Control''}, and \emph{``Resource Governance''}—in inconsistent or interchangeable ways. Manual inspection revealed that many policies labeled as \emph{``Security Governance''} were actually enforcing direct access restrictions (e.g., SSH access rules, API authorization, file-path permissions). These were therefore consolidated under a unified \emph{``Access Control''} category. 
This manual refinement was essential to correct hallucinations, resolve misinterpretations, and ensure conceptual coherence in the taxonomy~\cite{zhang2025llm}. Of the 111 pilot samples, 95 LLM-proposed labels were judged as conceptually acceptable with minor modifications, whereas 18 required substantive discussion and re-labelling.
Through this open coding process, we arrived at an initial taxonomy composed of 5 main categories and 15 sub-categories, grounded in both automated and expert insights.

\subsubsection*{\dcircle{4}Full Dataset Labeling} 
Once the pilot study was done, we moved into the second phase of our process by labelling the remaining samples of the dataset (70\%), denoted as \( R \). 
However, this time, we adopted a refined version of the LLM prompt, as we used the initial taxonomy, derived from the pilot study, as a guide to the LLM.
The prompt explicitly asked the LLM to classify each policy according to the predefined categories and sub-categories. 
If a policy did not fit within the provided taxonomy, the LLM was instructed to suggest a new label.
A notable outcome of this phase is that the taxonomy derived from the pilot sample proved fully stable when applied to the full dataset. No new top-level categories emerged during the full labelling step, indicating that the pilot-derived categories captured the principal governance intents present across tools. Likewise, at the sub-category level, all sub-categories observed in the full dataset had already been identified during the pilot phase. This confirms that the pilot taxonomy provided comprehensive coverage of the policy space and required no structural extensions when applied to the remaining samples.
In contrast to the pilot study, two researchers independently reviewed the outputs. 
Inter-rater agreement was assessed using Cohen's Kappa coefficient \cite{cohen1960coefficient} to evaluate classification consistency across two dimensions: taxonomy category and sub-category. The resulting agreement scores were as follows: \textit{0.843} for category, \textit{0.840} for sub-category. These values indicate strong inter-rater reliability. All disagreements were resolved through discussion until full consensus was reached. 


\subsubsection{\textbf{Results}}

\begin{figure}
    \centering    \includegraphics[width=0.8\textwidth]{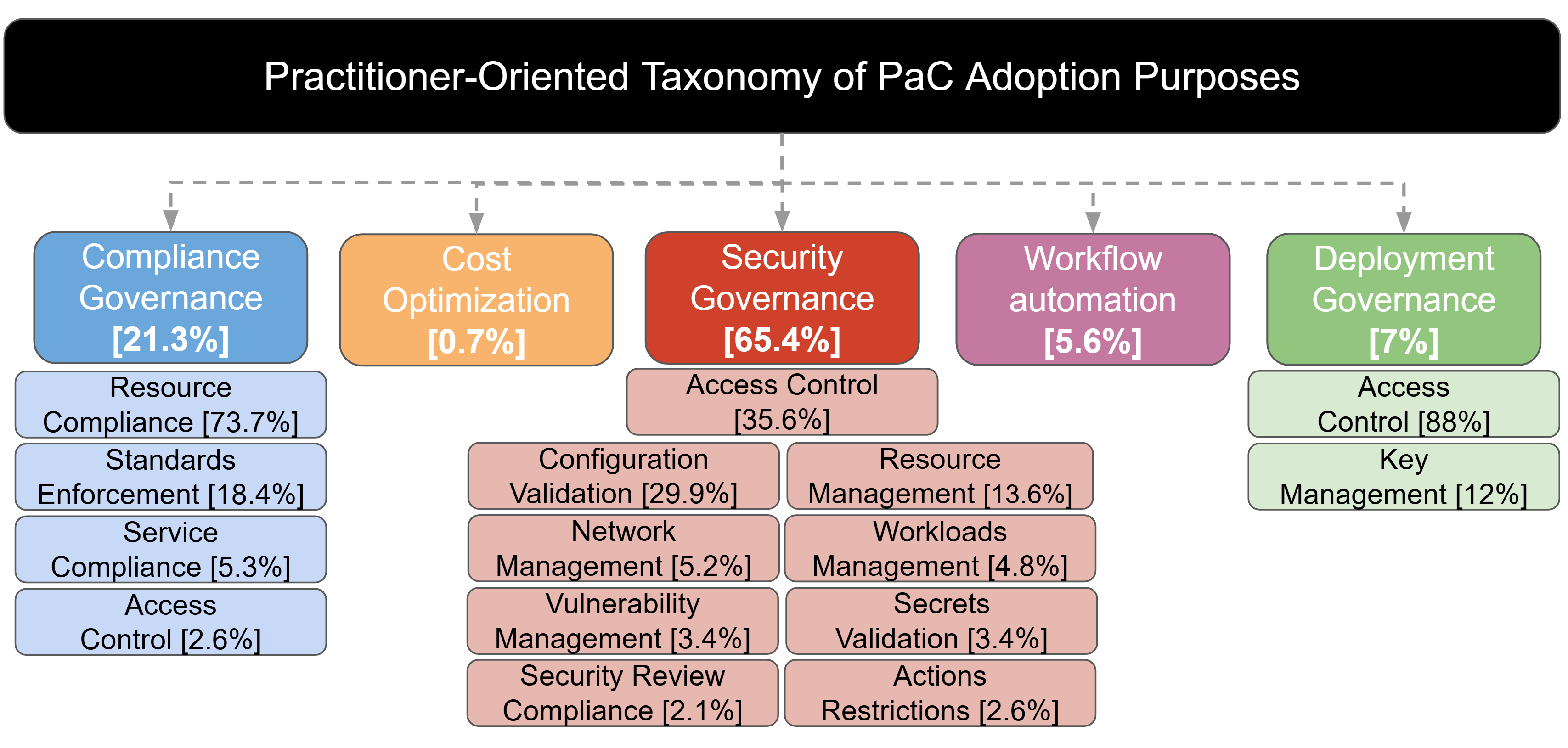}
    \caption{A Taxonomy of Policy-as-Code Usage and Adoption Patterns}
    \label{fig:taxonomy}
\end{figure}
\autoref{fig:taxonomy} presents the taxonomy of PaC usage, which consists of five main categories and fourteen sub-categories. During the taxonomy construction process, we were unable to classify 28 PaC files due to insufficient or incomplete code content.
Now, we describe each main category along with its associated sub-categories as follows.

\subsubsection*{\textbf{Security Governance}}
This category includes policies designed to protect software systems by defining and enforcing security-related rules. These rules ensure secure configurations, access restrictions, and appropriate behavior of infrastructure and services.
\begin{itemize}
    \item \textbf{Access Control:} This sub-category includes policies that restrict or grant access to specific resources based on roles, identity, or contextual factors. Examples include authorization policies for users, groups, or services that manage \textit{who} can access \textit{what} and under what conditions.
    \item \textbf{Configuration Validation:} These policies ensure that individual resources conform to specific security, structural, or operational constraints before being accepted or deployed. It focuses on validating fields such as volume types, base images, identity settings, and configuration correctness to prevent misconfigurations and enforce secure defaults.
    \item \textbf{Secrets Management:} These policies focus on securing sensitive data such as API tokens, credentials, and keys. Policies in this sub-category enforce best practices for handling secrets--e.g., ensuring secrets are not hard-coded or are retrieved securely from a secrets manager.
    \item \textbf{Network Management:} These policies regulate traffic flow, IP whitelisting, and connectivity rules within a system. They often include firewall rules, port restrictions, and control over ingress/egress configurations to reduce attack surfaces.
    \item \textbf{Resource Management:} These policies enforce constraints on the creation, usage, and allocation of system resources such as CPU, memory, containers, VMs, disk, clusters and Kubernetes pods. Typical policies ensure the creation, usage or allocation of secure system resources.
    \item \textbf{Security Review Compliance:} It contains policies that verify whether certain conditions indicating security review processes have been properly addressed. This ensures traceability and accountability during deployment and operational workflows.
    \item \textbf{Vulnerability Management:} These policies aim to detect and prevent the deployment of software with known vulnerabilities or missing updates. This may include policies enforcing up-to-date package versions or disallowing insecure dependencies.
    \item \textbf{Actions Restrictions:} This sub-category focuses on preventing the execution of specific actions or commands that are considered harmful, unauthorized, or non-compliant with security policies. For example, a policy might deny execution of any command listed in a predefined deny list within a cloud-native container environment.
    \item \textbf{Workloads Management:} This sub-category focuses on the configuration and enforcement of operational constraints for workloads. For example, policies may deny the deployment of workloads that automatically mount service account tokens to enhance runtime security.
\end{itemize}

\subsubsection*{\textbf{Compliance Governance}}
This taxonomy category ensures systems adhere to organizational policies, industry standards, and legal requirements. It covers internal best practices, external regulations, and service-specific constraints.
\begin{itemize}
    \item \textbf{Resource Compliance:} Policies that enforces structural or content rules on resource definitions to comply with organizational standards. These policies verify whether a resource instance meets defined compliance criteria--typically structural, security, or configuration requirements. These policies operate at the level of resource correctness, such as ensuring a Kubernetes Pod has a required label, a Terraform file is in valid format, or a YAML config is syntactically and semantically sound.
    \item \textbf{Service Compliance:} These policies ensure that services meet specific operational, security, or performance requirements--e.g., ensuring logging or monitoring is enabled for critical services.
    \item \textbf{Standards Enforcement:} This category enforces adherence of a particular technology to its best practice implementations, often to improve maintainability, reusability, and operational consistency. It includes policies that ensure consistent naming, approved module usage, tagging standards, and container optimization across systems. For example, a common policy requires the inclusion of \texttt{yum clean all} after \texttt{yum install} commands in Dockerfile \footnote{\url{https://shorturl.at/QnN19}} to ensure optimized image sizes and prevent unnecessary cache buildup. 
\end{itemize}

\subsubsection*{\textbf{Cost Optimization}}
This category includes policies aimed at reducing unnecessary cloud spending. Examples include enforcing resource quotas, limiting compute capacity, and identifying idle or underutilized resources.

\subsubsection*{\textbf{Workflow Automation}}
This category includes policies that trigger actions or workflows automatically, such as enforcing compliance during CI/CD, triggering audits, or provisioning of clouds infrastructure, resource allocation.

\subsubsection*{\textbf{Deployment Governance}}
This category includes policies that govern the secure, compliant, and auditable deployment of software systems. It ensures that deployments are performed by authorized entities, under predefined conditions, and in alignment with operational and security requirements--such as validating artifact integrity, enforcing configuration constraints, and controlling deployment permissions.
\begin{figure}
    \centering
    \includegraphics[width=0.5\textwidth]{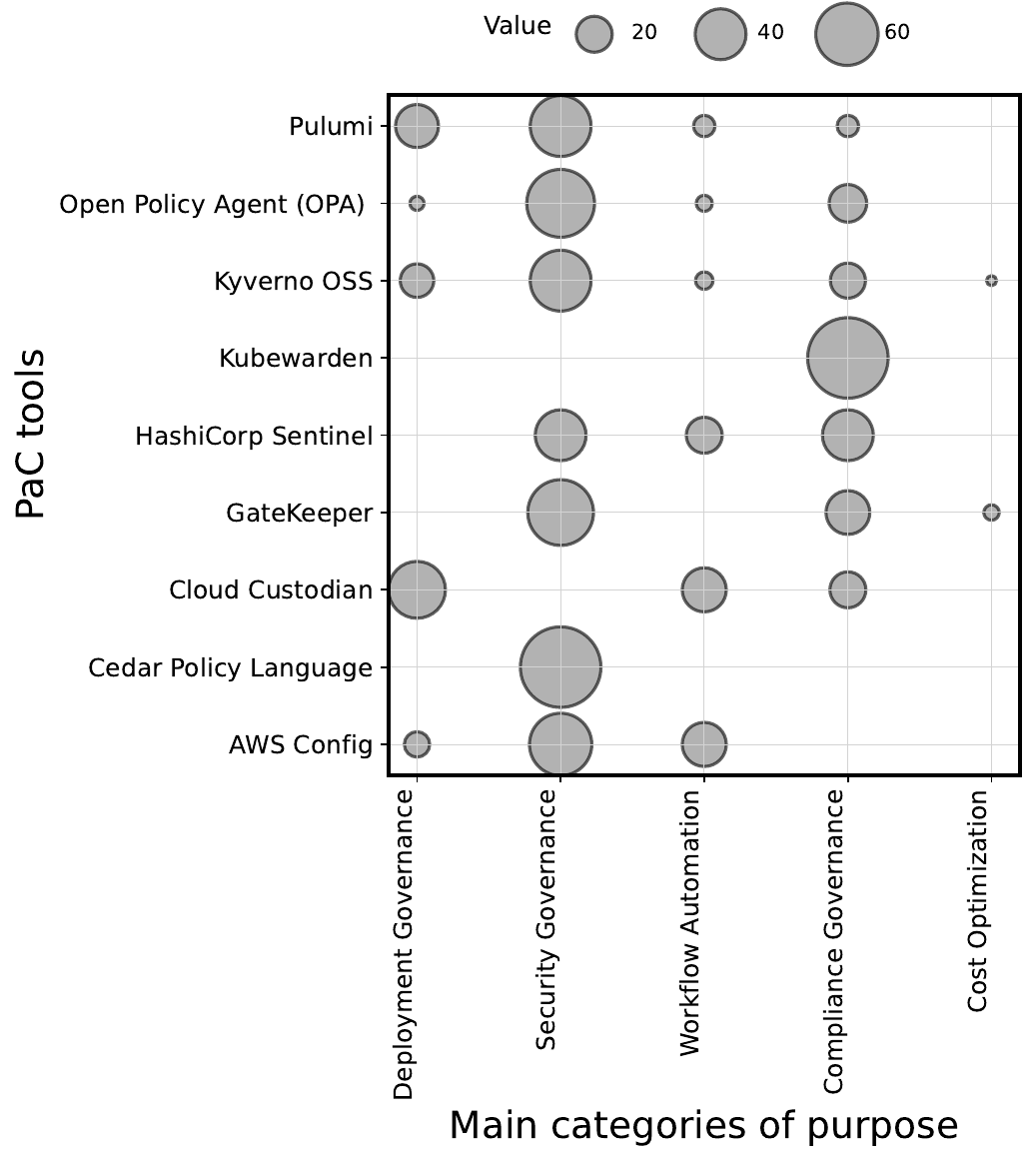}
    \caption{Mapping PaC Tools to Governance Categories in OSS (Bubble size = \% share)}
    \label{fig:pac_vs_purpose}
\end{figure}
\begin{itemize}
    \item \textbf{Key Management:} This sub-category focuses on the creation, monitoring, and lifecycle control of deployment-related keys. For example, a policy might trigger alerts when a new deploy key is created, ensuring visibility and accountability over sensitive credential usage during deployment workflows.
    \item \textbf{Access Control:} This sub-category focuses on regulating who can perform which deployment actions and under what conditions. Policies in this group ensure that only specific users or agents with verified identities and attributes are permitted to initiate deployment operations.
\end{itemize}
In complement to our taxonomy, \autoref{fig:pac_vs_purpose} offers a detailed breakdown of how different PaC tools are distributed across the five main governance categories. Each bubble's area is proportional to the percentage of policies within that tool that fall into the corresponding category 
(occurrences $\div$ total policies for that tool $\times$ 100). The legend markers (20/40/60) indicate representative percentage levels; the absence of a bubble denotes no observed policy for that tool--category pair. We clearly observe that \texttt{Security Governance}, and \texttt{Compliance Governance} dominate the landscape, appearing across nearly all tools. Tools like OPA, Kyverno, and Pulumi exhibit substantial focus on security policies while sharing other purposes. Conversely, \texttt{Cost Optimization} remains the least targeted category, with only a few scattered instances across tools like Kyverno OSS and GateKeeper. 
This figure shows the importance of tool-purpose alignment in real-world PaC adoption: while some tools offer broad applicability, others like kuberwarden and cedar are more specialized.

\begin{tcolorbox}[colback=gray!5!white,colframe=gray!75!black]
We identified five main categories of PaC usage, with Security Governance (65.4\%) and Compliance Governance (21.3\%) emerging as the most dominant purposes across the OSS ecosystem. While less frequent categories--such as Workflow Automation, Deployment Governance, and especially Cost Optimization--highlight emerging or context-specific adoption needs.
\end{tcolorbox}

\subsection{\textbf{RQ4: Are different PaC tools used together?}}
\subsubsection{\textbf{Motivation}} When addressing RQ1, we observed that OPA alone accounts for more than half of the PaC-related files, while RQ3 shows that some purposes are shared by multiple PaC tools. These findings raised the question of whether some PaC tools are commonly used in conjunction with other PaC tools. Understanding the co-usage of PaC tools can provide valuable insights for several stakeholders: (i) practitioners may benefit from knowledge about common tool combinations to inform their adoption strategies and integration workflows; (ii) researchers can further investigate the rationale behind tool combinations and their complementary use cases; and (iii) tool developers can enhance interoperability by designing features that facilitate integration with other commonly co-used PaC tools.

\subsubsection{\textbf{Approach}} 
To investigate whether different PaC tools are used together in the same projects, we performed a co-usage analysis across the 399 repositories. This analysis consisted of the following steps:

\subsubsection*{Binary Co-usage Matrix Construction} For each repository, we identified the set of tools used, based on the presence of corresponding policy files as defined in our identification patterns (see Section~\ref{sec:approach}). 
Next, we take this information and build a binary matrix where rows represent repositories and columns indicate the presence (1) or absence (0) of each PaC tool. 

\subsubsection*{Co-usage Frequency Calculation} Using the binary matrix $\mathbf{B}$ of size $n \times m$ (where $n$ is the number of repositories and $m$ the number of PaC tools), we computed a symmetric co-usage matrix $\mathbf{C}$ by multiplying the transpose of $\mathbf{B}$ with itself: $ \mathbf{C} = \mathbf{B}^\top\mathbf{B} $.
Each cell $C_{ij}$ in the resulting matrix represents the number of repositories where both tools $i$ and $j$ are used together.

\subsubsection*{Normalization \& Co-usage Intensity Analysis} To better understand the relative co-usage, we normalized the matrix to express values as a percentage of total repositories, allowing for intuitive identification of strong co-usage relationships between specific tools.
Finally, to complement the matrix co-usage, we also calculated how many distinct PaC tools were used per repository (e.g., using 1, 2, 3, or more tools).


\subsubsection{\textbf{Result}}
\begin{figure}
    \centering
    \includegraphics[width=0.6\textwidth]{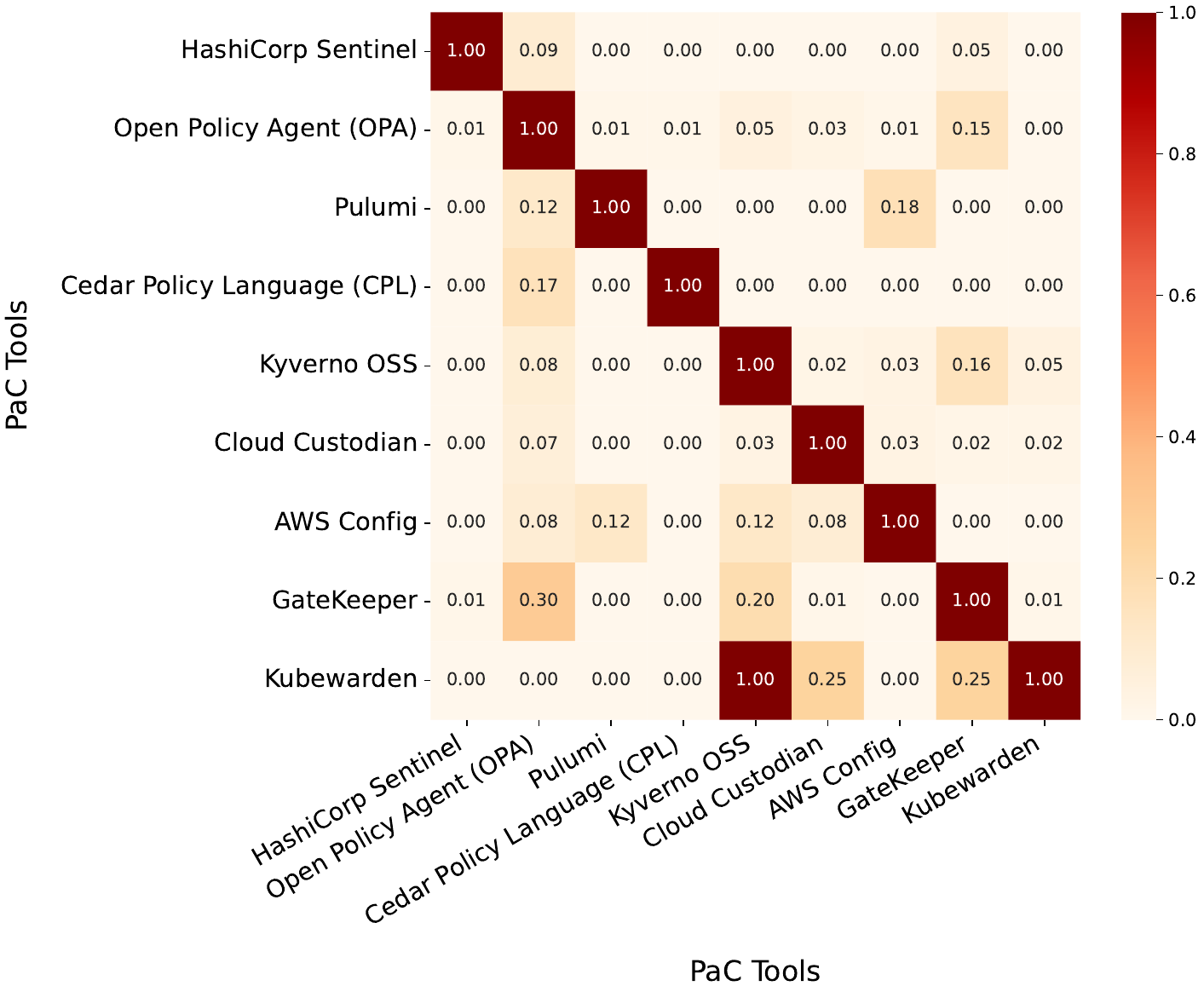}
    \caption{Co-usage matrix of PaC tools.}
    \label{fig:co_usage_matrix}
\end{figure}

Figure~\ref{fig:co_usage_matrix} presents the co-usage matrix between PaC tools, showing how frequently tools appear together across repositories. Most PaC tools are predominantly used in isolation, with co-usage values close to zero across most pairs. However, a few tools demonstrate notable integration patterns. In particular, OPA exhibits the highest co-usage with GateKeeper (30\%), followed by Cedar Policy Language (17\%) and Pulumi (12\%). Overall, OPA is used in combination with seven other PaC tools, except Kubewarden. This pattern can be explained by OPA's design as a general-purpose policy engine, which benefits from coupling with domain-specific tools that offer more context-aware policy abstractions, as highlighted in Section~\ref{sec:RQ3_result}. 
While the large majority of repositories (85.24\%) adopt only one PaC tool, 13.53\% of projects use two tools, and very few employ three tools simultaneously (1.23\%). Notably, only one project uses more than three PaC tools. 
This finding suggests that while PaC tools can be combined, they are most often adopted in isolation.



\begin{tcolorbox}[colback=gray!5!white,colframe=gray!75!black]
Most OSS projects (85.24\%) adopt only one PaC tool, showing a clear preference for isolated use. Multi-tool adoption is less common (14.76\%), most notably \textbf{OPA} with \textbf{GateKeeper}, though OPA also appears with seven other tools, suggesting emerging integration practices.  
\end{tcolorbox}

\section{DISCUSSION AND IMPLICATION}
\label{sec:discussion_implication}
This section presents the implications of our findings for the research community, practitioners, and tools developers.
\subsubsection*{\textbf{Researchers}}  
Our study offers a broad perspective on how practitioners use PaC tools in open-source projects. The results show that these tools are widely adopted in different project categories, particularly those serving as documentation (\autoref{tab:project_categories_summary}). Future research could investigate the factors shaping adoption, such as project maturity, community engagement, and organizational needs. Beyond documentation, PaC tools are most frequently used in traditional software systems, with limited adoption in Software 2.0 projects that embed AI or ML components. This underrepresentation suggests that the use of PaC in AI-enabled systems remains at an early stage, possibly because existing tools do not yet address the unique governance requirements of AI. This gap opens promising directions for research on how PaC frameworks could evolve to support trustworthiness, compliance, and lifecycle governance in intelligent systems, especially under emerging regulations such as the EU AI Act. 
Our methodological approach--combining LLM-assisted classification with expert validation--also illustrates both the potential and the limitations of using AI in empirical software engineering. This hybrid strategy highlights opportunities for advancing methods that integrate automated classification with human expertise, enabling the scalable analysis of large, heterogeneous code corpora while maintaining reliability and interpretability.  
Finally, our taxonomy provides a foundation for future extensions. As governance needs evolve in domains such as AI, cloud-native systems, and regulatory compliance, new categories and subcategories should be incorporated to capture emerging practices.  

\subsubsection*{\textbf{PaC Tool Developers}}  
Our results show that PaC tools are mainly used for security governance and compliance, with limited cases of advanced automation or multi-tool integration. Tool developers should therefore focus on improving the interoperability of PaC tools. Co-usage patterns--such as OPA frequently paired with GateKeeper--signal demand for complementary features that could be enabled through standardized APIs, modular architectures, or shared policy libraries to ease the combination of multiple PaC tools. The prominence of documentation-oriented projects further highlights the importance of accessible learning resources and well-designed examples as adoption drivers. Developers should also explore opportunities in emerging domains such as MLOps, where policies must extend beyond infrastructure governance to address fairness, accountability, and regulatory compliance. 

\subsubsection*{\textbf{Practitioners}}  
For practitioners, our study offers concrete insights into the adoption contexts and usage patterns of PaC tools. These findings can help them understand tool popularity and the diverse purposes for which PaC is applied, informing the design and implementation of policies in their own systems. Our results also show that PaC tools often align with specific purposes--for example, Kuberwarden, which is more specialized for compliance, and Cedar specializes in security governance. Recognizing these distinctions can guide practitioners in selecting tools that best match their deployment and governance needs, rather than treating all PaC tools as interchangeable.  

\section{RELATED WORKS}
\label{sec:related_work}
In this section, we review research that intersects with our investigation on the adoption and usage of PaC tools in software engineering.

Chinamanagonda et al.~\cite{chinamanagonda2021automating} highlight the growing reliance on automation to enforce cloud governance, showing how policy-driven frameworks improve compliance, security, and operational efficiency by reducing human error and enabling continuous monitoring.  
Xu et al.~\cite{xu2012aurasium} present Aurasium, a system that enforces security and privacy policies in Android applications by repackaging APKs with user-level sandboxing. Their approach demonstrates how static instrumentation and runtime monitoring enable fine-grained enforcement without OS modifications.  

Verdet et al.~\cite{verdet2025assessing} conduct an empirical study of security policies in Terraform scripts across AWS, Azure, and GCP. Using static analysis tools such as Checkov and Tfsec, they assess the prevalence and enforcement of 682 security policies. While focused on IaC, their findings complement our work by providing evidence on policy adoption in cloud infrastructure, whereas we broaden the scope to include PaC tools for access control, compliance, and validation (e.g., OPA, Kyverno).  
He et al.~\cite{he2006ensuring} introduce ReCAPS, a requirements-based method for integrating access control policy specification into the software development process. Their case study shows how inconsistencies between requirements, policies, and design artifacts can be systematically resolved, resonating with our interest in automating policy enforcement in OSS projects. 

Wei et al.~\cite{wei2025understanding} conducted a comprehensive multivocal literature review of the \textit{Everything-as-Code} (EaC) paradigm and proposed a conceptual taxonomy encompassing multiple dimensions, including \textit{Infrastructure-as-Code, Configuration-as-Code, and Policy-as-Code}, organized into six broader functional layers. Within this model, they emphasize security and compliance as central governance concerns across several PaC practices, which align with two of the dominant categories identified in our empirical taxonomy. Our study complements Wei et al.'s work by providing large-scale empirical evidence from open-source repositories, illustrating how these governance aspects manifest in practice. Furthermore, our analysis extends beyond the conceptual dimensions described by Wei et al. by uncovering additional, practice-driven categories, \textit{cost optimization, workflow automation, and deployment governance}, that emerge from real-world PaC adoption. 

Caracciolo et al~\cite{caracciolo2023policy} explore the implementation of compliance automation in cloud environments using the Policy-as-Code paradigm. The study develops a proof-of-concept integrating OPA Gatekeeper within a cloud-based DevSecOps pipeline to demonstrate how policies can be automatically enforced across cloud workloads. This work complements our research by providing a system-level evaluation of PaC effectiveness in enforcing compliance and security policies within practical DevOps workflows.

Together, these studies demonstrate how policy management and automation have been advanced across mobile, web, and cloud domains. Yet, to the best of our knowledge, no prior work has undertaken a large-scale empirical investigation of PaC adoption in open-source projects. Our study addresses this gap by providing a comprehensive analysis of PaC usage trends in real-world repositories.

\section{THREATS TO VALIDITY}
\label{sec:threats_to_validity}
Now, we discuss the thread on the validity of our work.

\subsubsection*{\textbf{Construct validity threats}} 
Threats to construct validity primarily concern errors in identifying PaC tools based on ad-hoc detection patterns. To reduce this risk, we systematically derived patterns from official tool documentation and made them publicly available in our replication package~\cite{replication}. Another concern is false positives, such as repositories including PaC files only as illustrative examples. We mitigated this risk in two ways. First, at the project selection stage, we excluded ``toy'' repositories using activity, popularity, and accessibility filters. Second, during taxonomy construction, we discarded files whose content was insufficient to determine an intended usage, thereby reducing the influence of non-representative artifacts on our findings.

\subsubsection*{\textbf{Internal validity threats}} 
We manually classified projects using PaC tools and analyzed a sample of PaC code to understand the types of projects adopting these tools and the purposes for which they are employed. Manual labeling, however, introduces risks to reliability, and even LLM-based classification can suffer from errors such as hallucinations. To mitigate these risks, we applied stratified sampling and a double-coding strategy: two researchers independently performed the labeling and evaluated inter-rater agreement, achieving substantial consistency (see Section \ref{sec:RQ3_result}), in line with established practices in software engineering research~\cite{openja2022studying, abbassi2025unveiling}. 
Despite these measures, we acknowledge that some practitioners' use of PaC tools may fall outside the scope of our taxonomy. 
Future surveys and qualitative studies would be valuable to validate and extend our findings.

Another possible threat arises from our decision to treat each PaC file as an atomic unit of analysis. This assumes that the dominant concern expressed in a file reflects its primary purpose. In practice, a file may encode multiple, overlapping concerns that are not fully captured by our categorization. We chose the file level because most PaC frameworks---such as Rego, Sentinel, or JSON/YAML-based engines---are built on declarative languages where the file constitutes the standard unit of authoring, execution, and distribution. This makes the file level the most natural lens through which to study developer practices. To reduce oversimplification, we designed our prompt and review process to capture both primary purposes and sub-purposes, and we applied normalization across heterogeneous outputs. Nevertheless, future work could explore finer-grained analyses, for instance at the level of individual rules or function calls, to better capture cases where a single file encodes multiple governance concerns.

\subsubsection*{\textbf{External validity threats}} 
Our dataset is restricted to open-source projects hosted on GitHub, which may not fully represent proprietary or enterprise contexts where PaC adoption could follow different practices and motivations. Moreover, our filtering criteria (e.g., minimum stars, forks, and recent activity) excluded smaller or inactive projects, potentially underrepresenting early experimentation or niche usage. While our findings offer valuable insights into adoption trends in the open-source ecosystem, caution is warranted when extrapolating them to industrial settings. Future work should examine custom in-house solutions and proprietary tools, which may exhibit distinct usage patterns.

\section{CONCLUSION AND FUTURE WORK}
\label{sec:conclusion}
This paper presented the first large-scale empirical study of PaC adoption in open-source projects. We examined nine prominent tools and showed that they are mainly used in traditional software systems, while beginning to gain traction in intelligent systems.  

Using LLM-assisted classification combined with systematic sampling and expert validation, we found that PaC tools are most commonly applied to security and compliance governance. Adoption is also shaped by resource environments, with Kubernetes-native tools and cloud-specific frameworks dominating their domains. Analysis of co-usage patterns revealed that, although OPA is occasionally combined with other tools, most projects adopt a single PaC tool in isolation.  

Future work will investigate underutilized features that may indicate usability challenges or opportunities for tool evolution, and quantitatively study factors influencing successful adoption across both open-source and proprietary contexts.

\section{ACKNOWLEDGMENT}
This work is supported by the DEEL project CRDPJ 537462-18 funded by the Natural
Sciences and Engineering Research Council of Canada (NSERC) and the Consortium for
Research and Innovation in Aerospace in Québec (CRIAQ), together with its industrial
partners Thales Canada inc, Bell Textron Canada Limited, CAE inc and Bombardier inc.

\bibliographystyle{elsarticle-num} 
\bibliography{cas-refs}

@article{foalem2025logging,
  title={Logging requirement for continuous auditing of responsible machine learning-based applications},
  author={Foalem, Patrick Loic and Silva, Leuson Da and Khomh, Foutse and Li, Heng and Merlo, Ettore},
  journal={Empirical Software Engineering},
  volume={30},
  number={3},
  pages={97},
  year={2025},
  publisher={Springer}
}

@misc{pac1,
  title         = {Policy as Code Tools \& Examples to Make Better Infrastructure Easier, Anywhere},
  organization  = {Puppet},
  year          = {2025},
  url           = {https://www.puppet.com/blog/policy-as-code},
  urldate       = {2025-03-01},
  note          = {Accessed March 2025}
}

@misc{pac2,
  title         = {GitHub REST API Documentation},
  organization  = {GitHub},
  year          = {2025},
  url           = {https://docs.github.com/en/rest?apiVersion=2022-11-28},
  urldate       = {2025-03-01},
  note          = {Accessed March 2025}
}

@inproceedings{openja2022studying,
  title={Studying the practices of deploying machine learning projects on docker},
  author={Openja, Moses and Majidi, Forough and Khomh, Foutse and Chembakottu, Bhagya and Li, Heng},
  booktitle={Proceedings of the 26th international conference on evaluation and assessment in software engineering},
  pages={190--200},
  year={2022}
}

@misc{pac3,
  title         = {Policy as Code},
  organization  = {Open Policy Agent Project},
  year          = {2025},
  url           = {https://shorturl.at/DQ7QX}, 
  urldate       = {2025-03-01},
  note          = {Accessed March 2025}
}

@misc{pac4,
  title         = {Everyone Loves Policy as Code, No One Wants to Write Rego},
  organization  = {Permit.io},
  year          = {2025},
  url           = {https://www.permit.io/blog/no-one-wants-to-write-rego},
  urldate       = {2025-03-01},
  note          = {Accessed March 2025}
}

@misc{pac5,
  title         = {Navigating the Challenges of Policy as Code in Azure},
  author        = {Hans Løken Sgard},
  year          = {2024},
  url           = {https://hlokensgard.no/2024/01/15/navigating-the-challenges-of-policy-as-code-in-azure/},
  urldate       = {2025-03-01},
  note          = {Accessed March 2025}
}

@misc{pac6,
  title         = {How Policy-as-Code Can Make Your Developers Infallible},
  organization  = {Axiomatics},
  year          = {2025},
  url           = {https://axiomatics.com/blog/how-policy-as-code-can-make-your-developers-infallible},
  urldate       = {2025-03-01},
  note          = {Accessed March 2025}
}

@article{chinamanagonda2021automating,
  title={Automating Cloud Governance-Organizations automating compliance and governance in the cloud},
  author={Chinamanagonda, Sandeep},
  year={2021},
journal={}
}

@inproceedings{xu2012aurasium,
  title={Aurasium: Practical policy enforcement for android applications},
  author={Xu, Rubin and Sa{\"\i}di, Hassen and Anderson, Ross},
  booktitle={21st USENIX Security Symposium (USENIX Security 12)},
  pages={539--552},
  year={2012}
}

@article{verdet2025assessing,
  title={Assessing the adoption of security policies by developers in terraform across different cloud providers},
  author={Verdet, Alexandre and Hamdaqa, Mohammad and Silva, Leuson Da and Khomh, Foutse},
  journal={Empirical Software Engineering},
  volume={30},
  number={3},
  pages={74},
  year={2025},
  publisher={Springer}
}

@inproceedings{he2006ensuring,
  title={Ensuring compliance between policies, requirements and software design: A case study},
  author={He, Qingfeng and Otto, Paul and Anton, Annie I and Jones, Laurie},
  booktitle={Fourth IEEE International Workshop on Information Assurance (IWIA'06)},
  pages={14--pp},
  year={2006},
  organization={IEEE}
}

@misc{pac10,
  title         = {Open Policy Agent (OPA)},
  organization  = {Open Policy Agent Project},
  year          = {2025},
  url           = {https://www.openpolicyagent.org/},
  urldate       = {2025-03-01},
  note          = {Accessed March 2025}
}

@misc{pac11,
  title         = {Kyverno},
  organization  = {Kyverno Project},
  year          = {2025},
  url           = {https://kyverno.io/},
  urldate       = {2025-03-01},
  note          = {Accessed March 2025}
}

@INPROCEEDINGS{7961663,
  author={Anthonysamy, Pauline and Rashid, Awais and Chitchyan, Ruzanna},
  booktitle={2017 IEEE/ACM 39th International Conference on Software Engineering: Software Engineering in Society Track (ICSE-SEIS)}, 
  title={Privacy Requirements: Present \& Future}, 
  year={2017},
  volume={},
  number={},
  pages={13-22},
  doi={10.1109/ICSE-SEIS.2017.3}}

@inproceedings{gonzalez2020state,
  title={The state of the ml-universe: 10 years of artificial intelligence \& machine learning software development on github},
  author={Gonzalez, Danielle and Zimmermann, Thomas and Nagappan, Nachiappan},
  booktitle={Proceedings of the 17th International conference on mining software repositories},
  pages={431--442},
  year={2020}
}

@inproceedings{majidi2022empirical,
  title={An empirical study on the usage of automated machine learning tools},
  author={Majidi, Forough and Openja, Moses and Khomh, Foutse and Li, Heng},
  booktitle={2022 IEEE International Conference on Software Maintenance and Evolution (ICSME)},
  pages={59--70},
  year={2022},
  organization={IEEE}
}

@misc{pac15,
  title         = {Open Policy Agent (OPA)},
  organization  = {Open Policy Agent Project},
  year          = {2025},
  url           = {https://www.openpolicyagent.org/},
  urldate       = {2025-06-01},
  note          = {Accessed June 2025}
}

@misc{pac16,
  title         = {HashiCorp Sentinel},
  organization  = {HashiCorp},
  year          = {2025},
  url           = {https://developer.hashicorp.com/sentinel/docs/concepts/policy-as-code},
  urldate       = {2025-06-01},
  note          = {Accessed June 2025}
}

@misc{pac17,
  title         = {Pulumi CrossGuard: Core Concepts},
  organization  = {Pulumi},
  year          = {2025},
  url           = {https://www.pulumi.com/docs/iac/using-pulumi/crossguard/core-concepts/},
  urldate       = {2025-06-01},
  note          = {Accessed June 2025}
}

@misc{pac18,
  title         = {Cedar Policy Language (CPL)},
  organization  = {Cedar Project},
  year          = {2025},
  url           = {https://docs.cedarpolicy.com/},
  urldate       = {2025-06-01},
  note          = {Accessed June 2025}
}

@misc{pac19,
  title         = {Kyverno OSS: Security Documentation},
  organization  = {Kyverno Project},
  year          = {2025},
  url           = {https://kyverno.io/docs/security/},
  urldate       = {2025-06-01},
  note          = {Accessed June 2025}
}

@misc{pac20,
  title         = {Cloud Custodian Documentation},
  organization  = {Cloud Custodian Project},
  year          = {2025},
  url           = {https://cloudcustodian.io/docs/},
  urldate       = {2025-06-01},
  note          = {Accessed June 2025}
}

@misc{pac21,
  title         = {AWS Config},
  organization  = {Amazon Web Services},
  year          = {2025},
  url           = {https://aws.amazon.com/config/},
  urldate       = {2025-06-01},
  note          = {Accessed June 2025}
}

@misc{pac22,
  title         = {OPA Gatekeeper Documentation},
  organization  = {Open Policy Agent Project},
  year          = {2025},
  url           = {https://open-policy-agent.github.io/gatekeeper/website/docs/},
  urldate       = {2025-06-01},
  note          = {Accessed June 2025}
}

@misc{pac23,
  title         = {Kubewarden},
  organization  = {Kubewarden Project},
  year          = {2025},
  url           = {https://www.kubewarden.io/},
  urldate       = {2025-06-01},
  note          = {Accessed June 2025}
}

@misc{pac24,
  title         = {What is Policy as Code (PaC) \& How Do You Implement It?},
  organization  = {Spacelift},
  year          = {2025},
  url           = {https://spacelift.io/blog/what-is-policy-as-code#policy-as-code-tools},
  urldate       = {2025-06-01},
  note          = {Accessed June 2025}
}

@article{foalem2024studying,
  title={Studying logging practice in machine learning-based applications},
  author={Foalem, Patrick Loic and Khomh, Foutse and Li, Heng},
  journal={Information and Software Technology},
  volume={170},
  pages={107450},
  year={2024},
  publisher={Elsevier}
}

@article{abbassi2025unveiling,
  title={Unveiling inefficiencies in llm-generated code: Toward a comprehensive taxonomy},
  author={Abbassi, Altaf Allah and Da Silva, Leuson and Nikanjam, Amin and Khomh, Foutse},
  journal={arXiv preprint arXiv:2503.06327},
  year={2025}
}

@article{seaman1999qualitative,
  title={Qualitative methods in empirical studies of software engineering},
  author={Seaman, Carolyn B.},
  journal={IEEE Transactions on software engineering},
  volume={25},
  number={4},
  pages={557--572},
  year={1999},
  publisher={IEEE}
}

@article{zhang2025llm,
  title={Llm hallucinations in practical code generation: Phenomena, mechanism, and mitigation},
  author={Zhang, Ziyao and Wang, Chong and Wang, Yanlin and Shi, Ensheng and Ma, Yuchi and Zhong, Wanjun and Chen, Jiachi and Mao, Mingzhi and Zheng, Zibin},
  journal={Proceedings of the ACM on Software Engineering},
  volume={2},
  number={ISSTA},
  pages={481--503},
  year={2025},
  publisher={ACM New York, NY, USA}
}

@article{cohen1960coefficient,
  title={A coefficient of agreement for nominal scales},
  author={Cohen, Jacob},
  journal={Educational and psychological measurement},
  volume={20},
  number={1},
  pages={37--46},
  year={1960},
  publisher={Sage Publications Sage CA: Thousand Oaks, CA}
}

@inproceedings{ouni2023empirical,
  title={An empirical study on continuous integration trends, topics and challenges in stack overflow},
  author={Ouni, Ali and Saidani, Islem and Alomar, Eman and Mkaouer, Mohamed Wiem},
  booktitle={Proceedings of the 27th International Conference on Evaluation and Assessment in Software Engineering},
  pages={141--151},
  year={2023}
}

@inproceedings{chen2020comprehensive,
  title={A comprehensive study on challenges in deploying deep learning based software},
  author={Chen, Zhenpeng and Cao, Yanbin and Liu, Yuanqiang and Wang, Haoyu and Xie, Tao and Liu, Xuanzhe},
  booktitle={Proceedings of the 28th ACM joint meeting on European software engineering conference and symposium on the foundations of software engineering},
  pages={750--762},
  year={2020}
}

@misc{replication,
  title        = "{Replication package}",
  howpublished = "\url{https://github.com/foalem/Pac-paper}",
  note         = "Retrieved August 2025",
author = ""
}

@inproceedings{anwar2018understanding,
  title={Understanding the hidden cost of software vulnerabilities: Measurements and predictions},
  author={Anwar, Afsah and Khormali, Aminollah and Nyang, DaeHun and Mohaisen, Aziz},
  booktitle={International Conference on Security and Privacy in Communication Systems},
  pages={377--395},
  year={2018},
  organization={Springer}
}

@book{taylor1994guidelines,
  title={Guidelines for evaluating and expressing the uncertainty of NIST measurement results},
  author={Taylor, Barry N and Kuyatt, Chris E and others},
  volume={1297},
  year={1994},
  publisher={US Department of Commerce, Technology Administration, National Institute of~…}
}

@article{act2024eu,
  title={The eu artificial intelligence act},
  author={Act, EU Artificial Intelligence},
  journal={European Union},
  year={2024}
}

@INPROCEEDINGS{11121731,
  author={Li, Hao and Bezemer, Cor-Paul and Hassan, Ahmed E.},
  booktitle={2025 IEEE/ACM 47th International Conference on Software Engineering: Software Engineering in Practice (ICSE-SEIP)}, 
  title={Software Engineering and Foundation Models: Insights from Industry Blogs Using a Jury of Foundation Models}, 
  year={2025},
  volume={},
  number={},
  pages={307-318},
  doi={10.1109/ICSE-SEIP66354.2025.00033}}

@article{wei2025understanding,
  title={Understanding Everything as Code: A Taxonomy and Conceptual Model},
  author={Wei, Haoran and Madhavji, Nazim and Steinbacher, John},
  journal={arXiv preprint arXiv:2507.05100},
  year={2025}
}

@misc{repo1,
  title        = "{Repo}",
  howpublished = "\url{https://github.com/aws-samples/cfn101-workshop}",
  note         = "Retrieved August 2025",
author = ""
}

@misc{repo2,
  title        = "{Repo}",
  howpublished = "\url{https://github.com/kubescape/regolibrary}",
  note         = "Retrieved August 2025",
author = ""
}

@misc{repo3,
  title        = "{Repo}",
  howpublished = "\url{https://github.com/cloudposse/atmos}",
  note         = "Retrieved August 2025",
author = ""
}

@misc{repo4,
  title        = "{Repo}",
  howpublished = "\url{https://github.com/aquasecurity/postee}",
  note         = "Retrieved August 2025",
author = ""
}

@misc{repo5,
  title        = "{Repo}",
  howpublished = "\url{https://github.com/deployKF/deployKF}",
  note         = "Retrieved August 2025",
author = ""
}

@misc{repo6,
  title        = "{Repo}",
  howpublished = "\url{https://github.com/DataBassGit/AssistAF}",
  note         = "Retrieved August 2025",
author = ""
}

@phdthesis{caracciolo2023policy,
  title={Policy as Code, how to automate cloud compliance verification with open-source tools},
  author={Caracciolo, Mattia},
  year={2023},
  school={Politecnico di Torino}
}

\end{document}